\documentclass[aps,prb,twocolumn,superscriptaddress,showpacs]{revtex4}
\usepackage{graphicx}
\usepackage{latexsym}
\usepackage{amssymb}
\usepackage{amsmath}
\usepackage{amsfonts}
\usepackage{bm}
\usepackage{multirow}
\usepackage{color}
\usepackage{comment}

\newcommand{\beq}{\begin{equation}}
\newcommand{\eeq}{\end{equation}}
\newcommand{\beqn}{\begin{eqnarray}}
\newcommand{\eeqn}{\end{eqnarray}}

\DeclareMathAlphabet{\mathbbold}{U}{bbold}{m}{n}

\makeatletter
\newcommand\xleftrightarrow[2][]{%
  \ext@arrow 9999{\longleftrightarrowfill@}{#1}{#2}}
\newcommand\longleftrightarrowfill@{%
\arrowfill@\leftarrow\relbar\rightarrow} \makeatother

\begin{document}

\title{Moir\'{e} Insulators viewed as the Surface of three dimensional Symmetry Protected Topological Phases}

\author{Chao-Ming Jian}
\affiliation{Kavli Institute of Theoretical Physics, Santa
Barbara, CA 93106, USA} \affiliation{ Station Q, Microsoft
Research, Santa Barbara, California 93106-6105, USA}

\author{Cenke Xu}
\affiliation{Department of Physics, University of California,
Santa Barbara, CA 93106, USA}

\begin{abstract}

Recently, correlated physics such as superconductivity and
insulator at commensurate fractional electron fillings has been
discovered in several different systems with Moir\'{e}
superlattice and narrow electron bands near charge neutrality.
Before we learn more experimental details and the accurate
microscopic models describing the insulators, some general
conclusions can already be made about these systems, simply based
on their symmetries and electron fillings. The insulator in the
Moir\'{e} superlattice is described by an effective spin-orbital
model with approximate higher symmetries than ordinary spin
systems. We demonstrate that both the insulators observed at the
1/2 and 1/4 fillings away from the charge neutrality can be viewed
as the boundary of a three dimensional bosonic symmetry protected
topological phase, and hence have the 't Hooft anomaly once the
spatial symmetries are viewed as internal symmetries.

\end{abstract}

\pacs{}

\maketitle

\section{Introduction}

The Lieb-Schultz-Mattis (LSM) theorem~\cite{LSM}, and its higher
dimensional generalizations~\cite{oshikawa,hastings} state that if
a quantum spin system defined on a lattice has odd number of
spin-1/2s per unit cell, then any local spin Hamiltonian which
preserves the spin and translation symmetry, cannot have a
featureless (gapped and nondegenerate) ground state. A system
protected by the LSM theorem is very similar to the
boundary of a symmetry protected topological (SPT)
phase~\cite{wenspt,wenspt2}: with certain symmetry, the boundary
of the SPT state cannot be a featureless gapped state. Thus in
recent years many works have made the connection between the LSM
theorem and related physics in the $d-$dimensional space to the
boundary of systems with one higher
dimension~\cite{oshikawa2,xulsm,maxlsm,ryulsm,xutriangle,oshikawa3,chenglsm},
and once the lattice symmetry is viewed as an internal onsite
symmetry, the LSM theorem can be interpreted as the consequence of
the 't Hooft anomaly at the boundary of the higher dimensional
parent SPT phase.

Recently surprising correlated physics has been discovered in
different systems with Moir\'{e} superlattice, such as
superconductivity and insulator at fractional
fillings~\cite{wangmoire,mag01,mag02,young2018}, which motivated a
series of active theoretical
studies~\cite{xuleon,senthil,kivelson,fu,vafek,phillips,phillips2,ma,yang,louk,fu2,fu3,ashvinyou,martin,zhang,bernevig,scalet,senthil2,senthil3,fczhang,subir,balents2,dai}
(a consensus of the nature of the observed insulating behavior has
not been reached, in the current work we assume these insulators
are Mott insulators, and hence are described by a low energy
effective spin-orbital model). These systems have narrow electron
band width near charge neutrality, hence the interaction effects
are effectively enhanced near charge neutrality.

In two systems that are microscopically rather different, $i.e.$
(1) the heterostructure of trilayer graphene (TLG) and hexagonal
boron nitride (hBN), and (2) twisted bilayer graphene, MIs are
observed at both 1/2 and 1/4 fillings away from the charge neutral
point~\cite{wangmoire,mag01,young2018} (some of the insulating
behaviors were observed under pressure); superconductivity has
also been observed in both systems doped away from the MI
phases~\cite{mag02,young2018}.~\footnote{The superconductivity in
the heterostructure of TLG/hBN was confirmed through private
communication with Feng Wang.} The similar behaviors of these two
systems suggest a universal description. For the twisted bilayer
graphene system (TBLG), though studies based on a two-orbital
electron model on an effective triangular lattice were
pursued~\cite{xuleon,kivelson}, concerns were raised because the
triangular lattice tight binding model does not capture the band
touching at the charge neutral point, which is away from the Fermi
surface~\cite{senthil,fu,vafek}. But it was believed that such
triangular lattice tight binding model is fully justified for the
TLG/hBN heterostructure~\cite{senthil}. The similarity of the
recently observed phenomena in these two systems then suggests
that an analogous model may also be sufficient to describe the
most interesting correlated physics observed in TBLG.

In this work we assume that the MIs at the 1/2 and 1/4 fillings
can be described by a model on the effective triangular lattice,
which is definitely the case (at least) for the TLG/hBN
heterostructure. The charge fluctuation is frozen in the MI, thus
the system effectively is described by a model of spin and orbital
degrees of freedom, where the two orbitals are physically the two
valleys in the original Brillouin zone of graphene. Because the
valley/orbital polarization in this system is approximately
conserved since large-momentum transfer is highly suppressed due
to the long wave length modulation of the background potential in
the Moir\'{e} structure, the system has at least one extra
$U(1)_v$ symmetry, which corresponds to rotating the electrons at
the two valleys with opposite phase angles. Also, as was pointed
out in Ref.~\onlinecite{senthil,ashvinyou}, the exchange
interaction between the two orbitals (valleys), which would lead
to an effective Hund's coupling, could be rather weak in this
system, since it involves the overlap between the wave functions
at the two valleys.

So the electron model of the system should at least have $U(2)
\times U(1)_v \times \mathcal{T}$ symmetry, where the $U(2)$
contains the ordinary charge $U(1)$ and spin $SU(2)$ symmetry, and
$\mathcal{T}$ is the time-reversal symmetry. If we further ignore
the Hund's coupling, the electron model should have even higher
internal symmetry $[U(2)_L \times U(2)_R] \rtimes \mathcal{T}$,
where $L$ and $R$ label the two valleys, and $\mathcal{T}$
interchanges the two valleys. When the charge degree of freedom is
frozen in the MI, the symmetry of the electron model will be
inherited by the effective spin-orbital model. More detailed
analysis of these symmetries will be given in the next few
sections. The detailed microscopic models for a weak MI are
usually rather complicated and difficult to analyze (efforts of
deriving such models were made in
Ref.~\onlinecite{senthil,senthil2}), then the symmetries and
electron fillings are the only information we have without more
microscopic information of the MIs, and they are the key
ingredients for the analysis of LSM and anomaly related physics.

\section{Mott insulator at $1/4$ filling}

\subsection{With Hund's coupling}
\label{Sec:With Hund's}
The MI at $1/4$ filling has exactly one electron per Moir\'{e}
superlattice site. Throughout the paper we define the symmetry of
the effective spin-orbital model of the MI phase as the symmetry
of the operators that create local excitations without changing
the filling on each site. Then the symmetry of the spin-orbital
model of the MI at $1/4$ filling is \beqn SO(3)_s \times U(1)_v
\times \mathcal{T}. \eeqn Here we assume that there is only one
$SO(3)_s$ spin symmetry, as it is already sufficient to guarantee
a LSM theorem in this system. On every site of the triangular
lattice, there is a four dimensional Hilbert space, which forms a
spin-1/2 projective representation under $SO(3)_s$, and also a
projective representation under $U(1)_v \times \mathcal{T}$. We
can denote this representation as $(1/2_s, 1/2_v)$. Let's clarify
the meaning of the projective representation $1/2_v$. Here, we
normalize the periodicity of $U(1)_v$ such that the minimal but
non-trivial $U(1)_v$ charge value carried by local operators in
the effective spin-orbital model (such as the operator that hops
one electron from one valley to another) is set to be $\pm 1$.
Under this periodicity, the states in the four-dimensional Hilbert
space on each site carry charge $\pm \frac{1}{2}$ under $U(1)_v$,
which can be viewed as a projective representation under $U(1)_v
\times \mathcal{T}$. On top of this, since the 4-fold spin-orbit
Hilbert space is comprised of electronic states with single
occupancy, the time-reversal symmetry $\mathcal{T}$ should square
to $-1$, a property the $(1/2_s,1/2_v)$ representation of $SO(3)_s
\times U(1)_v \times \mathcal{T}$ always carry in our definition.

Our goal is to interpret the system as the boundary of a three
dimensional SPT state, hence it has a 't Hooft anomaly, as was
discussed in other systems in
Ref.~\onlinecite{oshikawa2,xulsm,maxlsm,ryulsm,xutriangle,oshikawa3,chenglsm}.
The anomaly of a system can be analyzed in pretty much any state
of the system, due to the anomaly matching
condition~\cite{anomalymatching,anomalymatching2}. We will select
a $Z_2$ spin liquid that can in principle exist in this
spin-orbital system on the triangular lattice, and $Z_2$ spin
liquids can be naturally constructed at the boundary of three
dimensional SPT states~\cite{senthilashvin}.

To construct this $Z_2$ spin liquid, we first introduce a four
component complex bosonic ``spinon" $b_\alpha$ on every site of
the triangular moir\'{e} superlattice. $b_\alpha$ forms a
$(1/2_s,1/2_v)$ representation of the $SO(3)_s \times U(1)_v
\times \mathcal{T}$ symmetry. We then impose a local constraint
\beqn \sum_{\alpha = 1}^4 b^\dagger_{j,\alpha}b_{j,\alpha} = 1,
\eeqn on every site $j$. The local constraint above will lead to a
$U(1)$ gauge degree of freedom, namely $b_\alpha$ is not a gauge
invariant operator, it couples to a dynamical $U(1)$ gauge field.
Thus we expected an excitation with the $(1/2_s,1/2_v)$
representation to be a fractionalized excitation of the MI.

Then a $Z_2$ spin liquid, or a $Z_2$ topological order can be
constructed by condensing the pair of the $b_\alpha$ field, and
$b_\alpha$ is the $e$ particle of the $Z_2$ topological order. In
the $Z_2$ spin liquid, besides the spinon $b_\alpha$, there is
also a vison excitation, which is the $\pi-$flux of the original
$U(1)$ gauge field, and also the $m$ excitation of the $Z_2$
topological order. The dynamics of the vison is frustrated by the
spinon on every site, and its dynamics is described by a fully
frustrated quantum Ising model on the dual honeycomb lattice. The
vison Hamiltonian should be invariant under the spin-orbit
symmetry $SO(3)_s \times U(1)_v \times \mathcal{T}$. The fully
frustrated quantum Ising model should have the entire lattice
symmetry such as the $C_3$ rotation. Hence in its Brillouin zone
there are four symmetry protected minima as was previously studied
in Ref.~\onlinecite{sondhiz2gauge}, and the low energy modes of
the visons should be particle-hole conjugated by the time-reversal
symmetry $\mathcal{T}$, as time-reversal will reverse the momentum
of the low energy vison modes. The low energy dynamics of the
vison has a large emergent symmetry which protects the degeneracy
of the minima in the vison band structure, and if the vison
condenses, it will drive the system into a valence bond solid
(VBS) type of state that only spontaneously breaks the lattice
symmetry. Due to the emergent symmetry, the ground state manifold
(GMS) of the VBS state can be most conveniently embedded into a
group manifold $SO(3)_m$~\cite{sondhiz2gauge}.

Using the formalism developed in Ref.~\onlinecite{xuz2gauge}, the
$Z_2$ spin liquid mentioned above can be captured by a mutual
Chern-Simons (CS) theory: \beqn \mathcal{L} &=& \sum_{\alpha =
1}^4 |(\partial - i a)z_\alpha|^2 + \sum_{\beta = 1}^2 |(\partial
- ic)v_\beta|^2 \cr\cr &+& r_z |z_\alpha|^2 + r_v |v_\beta|^2 +
\frac{i}{\pi} a\wedge dc + \cdots \label{CS} \eeqn $z_\alpha$ is a
four-component complex boson field which carries a $(1/2_s,
1/2_v)$ representation under the full spin-orbital symmetry
$SO(3)_s \times U(1)_v \times \mathcal{T}$; the two-component
complex boson field $v_\beta$ carries a spinor representation of
the $SO(3)_m$ group; Besides the action on the boson field
$z_\alpha$, the time-reversal symmetry also transforms the vison
field $v_\beta$ and the gauge field $a$ and $c$ as $\mathcal{T}$:
$v_\beta \rightarrow v_\beta^*,~ (a_0,a_1,a_2)\rightarrow
(a_0,-a_1,-a_2),~ (c_0,c_1,c_2)\rightarrow (-c_0,c_1,c_2)$. One
can check that the mutual Chern-Simons (CS) theory Eq.\ref{CS} is
invariant under the full spin-orbital symmetry $SO(3)_s \times
U(1)_v \times \mathcal{T}$ and the $SO(3)_m$ group. When
$z_\alpha$ and $v_\beta$ are both gapped ($r_z, r_v
>0$), they are the $e$ and $m$ excitations of a symmetric $Z_2$
spin liquid on the triangular lattice, with a mutual semion
statistics enforced by the mutual CS term. The VBS phase mentioned
in the previous paragraph which corresponds to the condensate of
visons can be obtained by keeping $r_z > 0$, while bringing $r_v <
0$. Then after integrating out the gapped $b_\alpha$ field, the
vison field $v_\beta$ is coupled to a mutual CS field which is
equivalent to a $Z_2$ gauge field, and the condensate of $v_\beta$
has the ground state manifold $SO(3)_m$ which can be described by
three orthogonal gauge invariant vectors~\cite{senthilchubukov}:
\beqn v^\dagger \vec{\sigma} v, \ \ \ \mathrm{Re}[v^t \sigma^y
\vec{\sigma} v], \ \ \ \mathrm{and} \ \ \mathrm{Im}[v^t \sigma^y
\vec{\sigma} v]. \label{op}\eeqn

Eq.~\ref{CS} is sufficient for us to ``derive" the 't Hooft
anomaly. But let us first discuss the physical construction of the
bulk SPT phase in three dimensions, whose boundary is the same
$Z_2$ topological order given by Eq.~\ref{CS}. We will try to
construct a $3d$ SPT phase with $SO(3)_s \times U(1)_v \times
\mathcal{T} \times SO(3)_m$ onsite symmetry through the standard
``decorated defect"
procedure~\cite{senthilashvin,xuclass,xutriangle}:

(1) In the $3d$ bulk, we first consider an ordered phase whose
order parameter forms a ground state manifold $SO(3)_m$;

(2) Then in the ordered phase of the $SO(3)_m$ order parameter,
due to the fact that $\pi_1[SO(3)] = Z_2$, there is a $Z_2$ vortex
line topological excitation (defect);

(3) Then we decorate this $Z_2$ vortex line with the $1d$ SPT
phase with $SO(3)_s \times U(1)_v \times \mathcal{T}$ symmetry
whose boundary is a $(1/2_s, 1/2_v)$ representation;

(4) Eventually we restore all the symmetry in the bulk by
proliferating/condensing the vortex loops which have been
decorated with the $1d$ SPT phase.

To elaborate this construction, we need to review the $1d$ SPT
phases. There is a standard $1d$ Haldane SPT phase with $SO(3)_s$
symmetry~\cite{haldane1,haldane2} with $Z_2$ classification.
$U(1)_v \times \mathcal{T}$ SPT state in $1d$ has a $Z_2 \times
Z_2$ classification\cite{wenspt,wenspt2,wenSO}. There is one $1d$
$U(1)_v \times \mathcal{T}$ SPT state whose $0d$ boundary carries
half charge under $U(1)_v$, and also a Kramers doublet such that
the time-reversal action squares to $-1$.
It is the {\it product} of the two $1d$ SPT phases mentioned above
that we decorate into the vortex line of the $SO(3)_m$ order
parameter in the previous paragraph, which again has a $Z_2$
classification itself, namely two copies of this product $1d$ SPT
phases become a trivial phase. The $0d$ boundary of this $1d$ SPT
phase is precisely the $(1/2_s, 1/2_v)$ representation under the
$SO(3)_s \times U(1)_v \times \mathcal{T}$ symmetry. The $Z_2$
nature of the vortex line of the ground state manifold $SO(3)_m$
is perfectly compatible with the $Z_2$ classification of the $1d$
SPT phase decorated along the vortex lines.

Please note that in this construction, we will proliferate the
decorated closed vortex loops in the bulk, but we will not (in
fact we cannot) proliferate the ``termination" of a single vortex
line on the $2d$ boundary without breaking any symmetry, because
the termination of a single vortex line at the boundary
corresponds to the boundary of the $1d$ SPT phase, which carries a
projective representation of the spin-orbital symmetry, whose
condensation will lead to spontaneous breaking of the spin-orbital
symmetry. However, at the $2d$ boundary, we can disorder the
$SO(3)_m$ order parameter, while keeping a finite gap of the
vortex of the $SO(3)_m$ order parameter. Then the system enters a
$Z_2$ topological order whose $e$ particle is the gapped vortex of
the $SO(3)_m$ order parameter with a $(1/2_s, 1/2_v)$ projective
representation, and the $m$ particle is the ``fractionalized" $SO(3)_m$ order
parameter, which is precisely the $v_\beta$ field Eq.~\ref{CS},
and $v_\beta$ is connected to the $SO(3)_m$ order parameter
through Eq.~\ref{op}.

A similar $Z_2$ spin liquid as Eq.~\ref{CS} can be naturally
constructed for an ordinary spin-1/2 system on the triangular
lattice, whose $e$ and $m$ excitations carry spin-1/2
representations of $SO(3)_s$ and $SO(3)_m$ group
respectively~\cite{xuz2gauge}. In
Ref.~\onlinecite{lutriangle,xutriangle} it was shown that there is
a ``parent state" of this $Z_2$ spin liquid, which is an algebraic
spin liquid state described by the $N_f = 4$ QED$_3$ with four
flavors of (two-component) Dirac fermions coupled with a dynamical
$U(1)$ gauge field, and hence has an emergent $SU(4)$ flavor
symmetry~\footnote{More precisely the symmetry is $SO(6)\sim
SU(4)/Z_2$, because the $Z_4$ center of the $SU(4)$ flavor
symmetry is gauged, while the gauge neutral monopole of the gauge
field, which is also a physical operator, carries a vector
representation of $SO(6)$. }. The $SO(3)_s$ and $SO(3)_m$
symmetries are both subgroups of the $SU(4)$ flavor symmetry, and
a spin-1/2 system on the triangular lattice can be viewed as the
boundary of a $3d$ SPT phase with $SO(3)_s \times SO(3)_m$
symmetry. The $3d$ bulk SPT is constructed by decorating the
vortex line of the $SO(3)_m$ manifold with a $SO(3)_s$ Haldane
phase~\cite{xutriangle}.

\begin{figure}
\includegraphics[width=220pt]{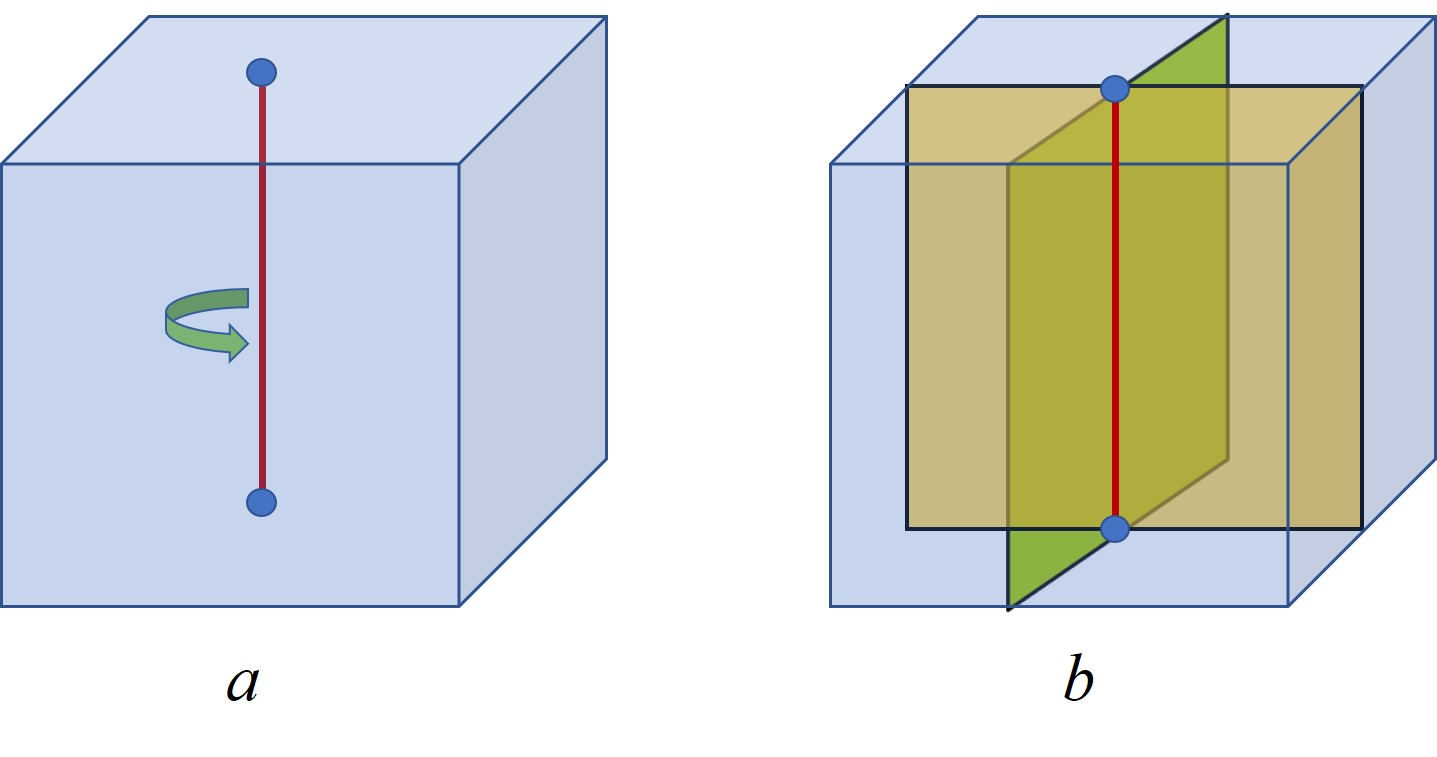}
\caption{($a$) The ``decorated defect" construction of the $3d$
bulk SPT state: for the MI with $1/4$ filling, we decorate a $1d$
SPT with $SO(3)_s \times U(1)_v \times \mathcal{T}$ symmetry in
every $Z_2$ vortex line of the $SO(3)_m$ order parameter manifold;
($b$) When the rotation symmetry of the triangular lattice is
broken, while the translation symmetry is preserved, the
translation symmetry becomes the $Z_2 \times Z_2$ subgroup of
$SO(3)_m$, and the vortex line is reduced to the intersection of
two $Z_2$ domain walls.} \label{spt}
\end{figure}

It is natural to see that in the parent QED$_3$ state the
translation symmetries of the lattice correspond to the $Z_2
\times Z_2$ subgroup of the $SO(3)_m$ group, $i.e.$ they are the
$\pi-$rotation around two orthogonal axes (more detail is given in
the appendix). In fact, as long as we preserve the translation
symmetry of the triangular lattice (but break the rotation
symmetry), it is already sufficient to guarantee a LSM theorem. In
this case, the system can still be viewed as the boundary of a
$3d$ SPT state, while now the $SO(3)_m$ vortex line becomes the
$1d$ intersection of the domain walls of the two $Z_2$ subgroups
(Fig.~\ref{spt}), and the bulk is still a nontrivial $3d$ SPT
phase once we decorating this domain wall intersection with a $1d$
SPT phase.

In our current case a similar QED$_3$ parent state can be
constructed. In order to do that, we need to introduce two types
of fermionic partons on each site. We introduce $f_s = (f_{s,1},
f_{s,2})^t$ that forms a spin-$\frac{1}{2}_s$ representation under
$SO(3)_s$, that is charge neutral under $U(1)_v$ and has a
time-reversal actions $\mathcal{T}$ that squares to $-1$. We also
introduce the parton $f_v = (f_{v,1}, f_{v,2})^t$ that is a
doublet consisting of modes with charge $\pm \frac{1}{2}$ under
$U(1)_v$. $f_{v,1}$ and $f_{v,2}$ are interchanged under the
$\mathcal{T}$ action such that the time-reversal action squares to
$+1$. $f_v$ transform trivially under $SO(3)_s$.
Then by imposing the constraint that $f_v$ and $f_s$ each has
exactly one fermion on each site $j$: \beqn \sum_{\alpha = 1}^2
f^\dagger_{j,s,\alpha} f_{j,s,\alpha} = 1, \ \ \ \sum_{\alpha =
1}^2 f^\dagger_{j,v,\alpha} f_{j,v,\alpha} = 1, \eeqn the Hilbert
space on each site still forms a $(1/2_s, 1/2_v)$ projective
representation of the spin-orbital group $SO(3)_s \times U(1)_v
\times \mathcal{T}$. Then $f_s$ and $f_v$ can each form a $N_f =
4$ QED$_3$ with two distinct dynamical $U(1)$ gauge fields, $i.e.$
the parent state of the $Z_2$ spin liquid Eq.~\ref{CS} is two
copies of $N_f = 4$ QED$_3$.

Let us discuss the connection between the two copies of $N_f = 4$
QED$_3$ and the aforementioned $Z_2$ spin liquid. We start with
the two copies of QED$_3$. In the $N_f=4$ QED$_3$ associated to
the $f_s$ parton, the low-energy Dirac fermions, denoted as
$\psi_s$, carry the same quantum number as the $f_s$ parton under
symmetry $SO(3)_s \times \mathcal{T}$. $\psi_s$ also forms a
doublet representation under $SO(3)_m$. Overall, the $\psi_s$
fermion forms a $N_f=4$ dimensional representation of the total
group $SO(3)_s \times \mathcal{T} \times SO(3)_m$. The $U(1)$
gauge field that couples to $\psi_s$ transforms in the same way as
the standard electromagnetism under $\mathcal{T}$. In this
QED$_3$, we can turn on a singlet pairing of the $\psi_s$ that is
invariant under the total group. This pairing higgs the $U(1)$
gauge group down to $Z_2$, which indicates the $Z_2$ topological
order in the Higgs phase. The fermion $\psi_s$ will naturally be
identified as the fermionic particle in the $Z_2$ topological
order. There are two types of (deconfined) $\pi$-fluxes that will
be identified as the ``electric" and ``magnetic" particles of the
$Z_2$ topological order \cite{xutriangle}. We denote them as $e_1$
and $m_1$ particles. They differ from each other by a fermion
$\psi_s$ and they have the semionic mutual statistics. The $m_1$
particle transforms trivially under the spin-orbital group
$SO(3)_s \times \mathcal{T}$ but forms a doublet representation
under $SO(3)_m$. The $e_1$ particle forms a spin-$\frac{1}{2}_s$
representation under $SO(3)_s$ such that the time-reversal
$\mathcal{T}$ squares to $-1$. $e_1$ is invariant under the
$U(1)_v$ and $SO(3)_m$ actions.

In the $N_f=4$ QED$_3$ associated to the $f_v$ parton, the
low-energy Dirac fermions, denoted as $\psi_v$, carry the same
quantum number as the $f_v$ parton under the spin-orbital symmetry
group $U(1)_v \times \mathcal{T}$. $\psi_v$ also forms a doublet
representation under $SO(3)_m$. Overall, the $\psi_v$ fermion
forms a $N_f=4$ dimensional representation of the total group
$U(1)_v \times \mathcal{T}\times SO(3)_m$. Similarly, we can turn
on the fermion pairing that is a singlet under the total group to
deform this QED$_3$ into another $Z_2$ topological order. Again,
the fermionic particle of the $Z_2$ topological order is naturally
identified with $\psi_v$. The two types of (deconfined)
$\pi$-fluxes can be identified as the ``electric" particle $e_2$
and ``magnetic" particle $m_2$ of the $Z_2$ topological order
\cite{xutriangle}. They differ from each other by a fermion
$\psi_v$ and they have mutual semionic statistics. The $m_2$
particle transforms trivially under the spin-orbital group $U(1)_v
\times \mathcal{T}$ but again forms a doublet representation under
$SO(3)_m$. The $e_2$ particle forms a doublet with $\pm
\frac{1}{2}$ $U(1)_v$ charges. The two components of the $e_2$
doublet are interchanged by the time-reversal symmetry
$\mathcal{T}$ such that the time-reversal symmetry square to $1$
on $e_2$. $e_2$ is invariant under the $SO(3)_s$ and $SO(3)_m$
actions.

Now we have constructed two copies of $Z_2$ topological order with
the quasi-particle contents $\{1,e_1,m_1,\psi_s\}$ and
$\{1,e_2,m_2,\psi_v\}$ where $1$ stands for the trivial particle.
Since $m_1$ and $m_2$ are both doublets under the $SO(3)_m$ group,
we can further condense the pair $m_1 m_2$ which is a singlet
under $SO(3)_m$ without breaking any symmetry.  The resulting
topological order is a single copy of $Z_2$ topological order
whose electric particle can be identified as $e_1 e_2$ and
magnetic particle as $m_1$ (which is equivalent to $m_2$). One can
immediately check that the deconfined particle $e_1e_2$ (which has
a trivial braiding statistics with the condensed bound state $m_1
m_2$) has the same statistics and symmetry quantum numbers as the
boson field $z_\alpha$ in the $Z_2$ spin liquid phase of Eq.
\ref{CS}. Also, the particle $m_1$ has the same statistics and
symmetry quantum numbers as the vison field $v_\beta$ in the $Z_2$
spin liquid phase of Eq. \ref{CS}. Now, we can conclude that we
have constructed the $Z_2$ spin liquid phase of the mutual CS
theory Eq. \ref{CS} from the two copies of $N_f=4$ QED$_3$.

As we discussed, this $Z_2$ spin liquid should be the boundary
state of a $3d$ SPT phase with $SO(3)_s \times U(1)_v \times
\mathcal{T} \times SO(3)_m$ that can be described using the
``decorated-defect" construction. From the construction of this
$3d$ SPT, we can directly write down its topological response
action \beqn \mathcal{S} = \pi \int (w_2[\mathcal{A}_s] +
\frac{1}{2\pi} dA_v+ w_1^2[TM]  )\cup w_2[\mathcal{A}_m] \eeqn
under the background $SO(3)_s$ gauge field $\mathcal{A}_s$,
$SO(3)_m$ gauge field $\mathcal{A}_m$, $U(1)_v$ gauge field $A_v$
and tangent bundle $TM$ of spacetime manifold. Here, $w_{1,2}$
stand for the first and the second Stiefel-Whitney classes. In
this topological response action, the $w_2 [\mathcal{A}_m]$ part
physically measures the $SO(3)_m$ flux.
The $w_2[\mathcal{A}_s]$ term captures the topological response of
the $1d$ $SO(3)_s$ Haldane chain. The $dA_v$ term describes a $1d$
SPT with a half $U(1)_v$ charge, which is protected by
$\mathcal{T}$, on its boundary. The $w_1^2[TM]$ term characterizes
a $1d$ SPT with a Kramers doublet (such that the time-reversal action
squares to $-1$) on its boundary. This topological response
implies non-trivial 't Hooft anomaly on the boundary state of this
$3d$ SPT, which in this case can be further viewed as an
implication of the LSM theorem in the effective spin-orbital model
on the Moir\'{e} superlattice site with filling $\frac{1}{4}$.
From the topological action, we see that the 't Hooft anomaly (and
hence the LSM theorem) should exist even if we consider only the
$SO(3)_s \times SO(3)_m$ symmetry or only the symmetry
$\mathcal{T}\times SO(3)_m$, where $SO(3)_m$ essentially
represents the space-group symmetry that at least includes the
lattice translation symmetries. This means that even if we break
the $SO(3)_s$ and $U(1)_v$, as long as $\mathcal{T}$ and
translation is still preserved, there is still a LSM theorem for
this system, and the system can still be viewed as the boundary of
a nontrivial $3d$ SPT state.

\subsection{Without Hund's coupling}

Now let us consider the case where the Hund's coupling is ignored,
which is justified in some limit since as we argued before in this
system the Hund's coupling is supposed to be weak. In this case
operators that create local excitations form a linear
representation of $SO(4)_s = [SU(2)_L \times SU(2)_R]/Z_2$, where
$SU(2)_L$ and $SU(2)_R$ are spin symmetries on the left and right
valleys respectively. Hence in this case the symmetry of this MI
is just $SO(4)_s \rtimes \mathcal{T} \times U(1)_v$. The $\rtimes$
symbol stems from the fact that time-reversal interchanges the two
$SU(2)$ subgroups, hence time-reversal acts like an improper
rotation.

Every site of the Moir\'{e} lattice carries a Dirac spinor
projective representation of the $SO(4)_s$ symmetry group, $i.e.$
a $(1/2,0) \oplus (0, 1/2)$ representation. A $Z_2$ topological
order can still be constructed in this system like Eq.~\ref{CS},
while in this case $b_\alpha$ is a Dirac spinor of $SO(4)_s$, and
also a projective representation of $U(1)_v \times \mathcal{T}$.
In fact, we can also embed the $SO(4)_s$ and $U(1)_v$ into a
$SO(6)_s$ group, and the four dimensional Hilbert space forms a
spinor of the enlarged $SO(6)_s$ symmetry.

The construction of the $3d$ bulk SPT phase is similar as before.
We can decorate the $Z_2$ vortex line of the $SO(3)_m$ order
parameter manifold with a $1d$ SPT phase whose $1d$ bulk has a
vector representation of $SO(4)_s$ and also a linear
representation of $U(1)_v \times \mathcal{T}$, while its $0d$
boundary has a Dirac spinor of $SO(4)_s$ and a projective
representation of $U(1)_v \times \mathcal{T}$. This $1d$ SPT phase
itself still has a $Z_2$ classification, which is consistent with
the $Z_2$ vortex line of the $SO(3)_m$ order parameter manifold.
Then the natural boundary of this $3d$ SPT state is still the
$Z_2$ topological order described above.

\section{Mott insulator at $1/2$ filling}

Weak MI behavior was also discovered at $1/2$ filling doped away
from charge neutrality, in both TLG/hBN heterostructure and
twisted bilayer graphene~\cite{wangmoire,mag01,young2018}. The
$1/2$ filling means that there are two extra electrons/holes per
unit cell away from charge neutrality in the Moir\'{e}
superlattice. If we use a triangular lattice model to describe
these systems, then for the twisted bilayer graphene, the symmetry
of the system guarantees that the nearest neighbor hopping of the
electrons has a $U(4)$ symmetry, while the $U(4)$ to $U(2)_L
\times U(2)_R$ symmetry breaking will happen only in second
neighbor hopping, as the symmetry permits a valley dependent
imaginary hopping between second neighbor sites; for the TLG/hBN
heterostructure, the valley dependent imaginary hopping is
permitted even between the nearest neighbor sites~\cite{senthil}.

Let us first start with the $U(4)$ limit of the systems. When the
charge degree of freedom is completely frozen, the system will be
described by a $SU(4)\sim SO(6)$ spin model with a six component
vector representation on every site. Under the $U(4)$ to $U(2)_L
\times U(2)_R$ symmetry breaking, the six states on every site
will be split into a four component $SO(4) \sim [SU(2)_L \times
SU(2)_R]/Z_2$ vector, and two degenerate $SO(4)$ singlet states.
Thus there are two possible scenarios: the microscopic Hamiltonian
either prefers the four component $SO(4)$ vector states, or the
two component $SO(4)$ singlet states. In the following we will
study each scenario separately. If the Hund's coupling is
considered, then at least the former case has a spin-1 left on
every site, which no longer has a LSM theorem. In principle one
can construct a featureless spin state by splitting each spin-1
into three spin-1 objects, and each of these ``fractionalized"
spin-1 object forming a Haldane chain along one of the three
directions of the triangular lattice. Thus in this section we will
ignore the Hund's coupling, $i.e.$ we assume there is no exchange
interaction between the two valleys.

\subsection{$SO(4)$ vector state on each site}

Let us first consider the case where the microscopic Hamiltonian
prefers to have a four-component vector states on each site of the
triangular moir\'{e} superlattice. In fact, we will consider the
limit where the two other states of the $SO(6)$ vector is
completely projected out. Physically, this case implies that the
microscopic Hamiltonian favors to have one and exact one electron
each valley (orbital) on every site.

A more careful analysis of the symmetry of the MI is necessary. It
is crucial to distinguish the symmetry of the electron system, and
the symmetry of the low energy effective spin model that describes
the MI. As we mentioned before, the electron symmetry is $[U(2)_L
\times U(2)_R] \rtimes \mathcal{T}$. However, the symmetry of the
spin model should be identified as the symmetry carried by the
creation operators of local spin excitations. For example, for an
ordinary spin system, whether it has integer or half-integer spin
in each unit cell, the local spin excitations always carry integer
spins, thus the local spin creation operators always have $SO(3)$
spin symmetry, and a spin-1/2 excitation can only be
``fractionalized". Using this perspective, under the assumption
that there is always one and precisely one electron on each
valley, the precise symmetry of the allowed local operators which
create spin excitations is \beqn [SO(3)_L \otimes SO(3)_R]\rtimes
\mathcal{T} = pSO(4)_s \rtimes \mathcal{T}. \label{sym} \eeqn This
symmetry group needs some explanation. Since we have projected the
onsite Hilbert space on one electron on each valley, an allowed
local excitation cannot mix the two valleys, $i.e.$ it cannot move
an electron from one valley to another. So an local spin operator
will take the form $c^\dagger_L \vec{\sigma} c_L$, or $c^\dagger_R
\vec{\sigma} c_R$, both form linear representations of $SO(3)_L
\otimes SO(3)_R$.

$pSO(4)$ is the $SO(4)$ group mod out its $Z_2$ center. Let us
recall that $SO(4) = [SU(2) \otimes SU(2)]/Z_2 $, where the $Z_2$
in the ``denominator" is the common $Z_2$ center of the two
$SU(2)$ subgroups, $i.e.$ the simultaneous $2\pi$ rotation of both
$SU(2)$ becomes the identity element in $SO(4)$. In $pSO(4)$ we
need to mod out another $Z_2$ subgroup, because as we discussed
above, the local spin excitation of the system must carry linear
representation of both $SO(3)_L$ and $SO(3)_R$, while a vector
representation of the original $U(2)_L \times U(2)_R$ electron
symmetry is now a projective representation of the spin symmetry
$pSO(4)$. Hence a four component $SO(4)$ vector spin excitation
can only be a ``fractionalized" excitation of the spin system, and
can only exist in a ``spin liquid".

The time-reversal symmetry $\mathcal{T}$ exchanges the two
valleys. 
Since the effective spin model Hilbert space on each site consists
of states with two electrons each site, the square of the
$\mathcal{T}$ action on the effective spin Hilbert space is $+1$
(instead of $-1$ as in the $\frac{1}{4}$ filling case). The
$U(1)_v$ acts completely trivially both on the operators and the
spin states in this effective spin model. Hence, we don't need to
include $U(1)_v$ in the discussion that follows.

As before, in order to expose the anomaly of the system, we can
construct a similar $Z_2$ spin liquid as Eq.~\ref{CS}. We first
define a four component complex bosonic ``spinon" $b_\alpha$ on
every site of the triangular moir\'{e} superlattice, $b_\alpha$
forms a $(1/2,1/2)$ representation under the $U(2)_L \times
U(2)_R$ electron symmetry, or a vector (projective) representation
of the $pSO(4)_s$ spin symmetry. The time-reversal symmetry
$\mathcal{T}$ squares to 1 on the boson field $b_\alpha$. We then
impose a local constraint $\sum_{\alpha = 1}^4
b^\dagger_{j,\alpha}b_{j,\alpha} = 1$, on every site $j$. Then a
$Z_2$ spin liquid, or a $Z_2$ topological order can be constructed
by condensing the pair of the $b_\alpha$ field, and $b_\alpha$ is
the $e$ particle of the $Z_2$ topological order. The vison field
$v_\beta$ still forms a spin-1/2 representation of the $SO(3)_m$
group.

This $Z_2$ spin liquid can again be viewed as the boundary of a
$3d$ SPT phase. To elaborate this bulk construction, we need to
review the $1d$ SPT phases with the $pSO(4)_s = SO(3)_L \times
SO(3)_R$ symmetry. These $1d$ SPT phases have classification $Z_2
\times Z_2$~\cite{psun}. The two ``root" states are basically the
Haldane phase of $SO(3)_L$ and $SO(3)_R$ respectively, $i.e.$ its
boundary carries either $(1/2, 0)$ or $(0, 1/2)$ projective
representation of $pSO(4)_s$. Then the SPT phase with a vector
representation $(1/2, 1/2)$ at the boundary is a product of both
root states, and itself has a $Z_2$ classification, namely two
copies of this SPT phases will trivialize themselves. It is this
product SPT phase that we decorate into the vortex line of a
$SO(3)_m$ order parameter in the bulk, and this product SPT phase
preserves the $Z_2$ improper rotation of the $pSO(4)_s$ symmetry
that exchanges $SO(3)_L$ and $SO(3)_R$, thus it also preserves the
time-reversal which acts as the improper rotation. Here, notice
that, unlike the $\frac{1}{4}$ filling case, we do NOT decorate
the $SO(3)_m$ vortex line by the $1d$ SPT state with a
time-reversal Kramers doublet on its boundary. This is because the
time-reversal action squares to $-1$ on a Kramers doublet, which
is incompatible with the representation of the boson field
$b_\alpha$ or the effective spin Hilbert space on each site.

From the decorated-defect construction of this $3d$ SPT, we can
write down its topological response action \beqn \mathcal{S} = \pi
\int (w_2[\mathcal{A}_L] + w_2[\mathcal{A}_R]) \cup
w_2[\mathcal{A}_m] \eeqn under the background $SO(3)_L$ gauge
field $\mathcal{A}_L$, $SO(3)_R$ gauge field $\mathcal{A}_R$ and
$SO(3)_m$ gauge field $\mathcal{A}_m$. This topological response
action indicates the non-trivial 't Hooft anomaly on the boundary
of this $3d$ SPT phase which can be further interpreted as the LSM
theorem on the Moir\'{e} superlattice site with filling
$\frac{1}{2}$ (under the $pSO(4)_s \rtimes \mathcal{T} \times
SO(3)_m$ symmetry). If we further turn on the Hund's coupling in
this effective model, the $SO(3)_L\times SO(3)_R$ symmetry is
broken down to its diagonal subgroup. In the topological response
theory, it amounts to setting $\mathcal{A}_L =\mathcal{A}_R$.
Since $w_2$ is a cohomology class with $Z_2$ coefficient, the
topological response theory becomes trivial when $\mathcal{A}_L
=\mathcal{A}_R$. Hence, the LSM theorem does not exist when the
Hund's rule coupling is turned on, which we expected before.

\subsection{$SO(4)_s$ singlet state on each site}

Now let us consider the case where the microscopic onsite
Hamiltonian strongly prefers two electrons on either the left or
right valley, hence it is a singlet under either $SU(2)_L$ or
$SU(2)_R$ electron symmetry. We will use $\tau^z = \pm 1$ to label
the two states with fully occupied left and right valleys
respectively, and the $U(1)_v$ transformation rotates the left and
right valleys by an opposite phase angle. The $U(1)_v$ symmetry
corresponds to the conservation of electron number on each valley
separately, which is a justified emergent symmetry at low energy.

Under time-reversal $\mathcal{T}$, $\tau^z$ changes its sign, thus
the symmetry in the two dimensional spin singlet Hilbert space is
$U(1)_v \times \mathcal{T}$. Here, we normalize the periodicity of
the $U(1)_v$ group such that the a local operator of this low
energy effective model (for example the operator that hops two
electrons on the left valley to the right valley) always carry
integer charge under $U(1)_v$. Under this normalization, the two
states on each site of this effective models carry $\pm
\frac{1}{2}$ $U(1)_v$ charge. Under $\mathcal{T}$, the two states
on each site are exchanged. This time-reversal action on the
quantum states squares to $1$. A $Z_2$ topological order can still
be constructed in this case, which is also described by
Eq.~\ref{CS}, the only difference is that $\alpha = 1,2$ in the
first term of the Lagrangian, and $z_\alpha$ transforms as a
projective representation of $U(1)_v \times \mathcal{T}$. The
dynamics of the vison is unchanged from Eq.~\ref{CS}.

This system again has a LSM theorem, $i.e.$ it cannot have fully
gapped nondegenerate ground state without breaking the $U(1)_v
\times \mathcal{T}$ symmetry. We can also map this system to the
boundary of a $3d$ SPT phase. The construction is similar to the
previous subsection: we consider the ordered phase in the $3d$
bulk with ground state manifold $SO(3)_m$; then we decorate this
$Z_2$ vortex line with the $1d$ SPT phase with $U(1)_v \times
\mathcal{T}$ symmetry whose boundary is a projective
representation of $U(1)_v \times \mathcal{T}$; eventually we
restore the symmetry in the bulk by proliferating the vortex
loops. From the decorated-defect construction of this $3d$ SPT, we
can write down its topological response action \beqn \mathcal{S} =
\pi \int \frac{1}{2\pi} A_v \cup w_2[\mathcal{A}_m] \eeqn under
the background $U(1)_v$ gauge field $A_v$ and $SO(3)_m$ gauge
field $\mathcal{A}_m$. The coefficient $\pi$ in the topological
response theory is protected by $\mathcal{T}$. Similar to the
previous discussions, this topological response action is tied to
the non-trivial LSM theorem.

\section{Discussion}

Given the novel effective symmetry groups of the moir\'{e} systems
discovered recently, we have analyzed possible spin-orbital states
of the weak Mott insulators observed at both 1/2 and 1/4 fillings
away from the charge neutrality. Due to the weak charge gap of the
MIs, the detailed effective spin-orbital Hamiltonian for these MIs
is expected to be rather complicated, thus in our paper we focused
on general analysis that only relies on the universal information
such as symmetry and electron filling. We mapped the MIs to the
boundary of $3d$ SPT phases, and demonstrated the 't Hooft anomaly
which is associated with the LSM theorem.

The two different moir\'{e} systems (TLG/hBN and TBLG) have
different detailed lattice symmetries such as reflection,
inversion, etc. Thus in our paper we focused on the common
symmetries shared by both systems. It is known that the 't Hooft
anomaly at the boundary of a $3d$ SPT state is usually related to
the deconfined quantum critical
point~\cite{deconfine1,deconfine2,senthilashvin,SO5,xulsm,maxlsm,xutriangle},
and the novel symmetries in the moir\'{e} systems may support new
types of deconfined quantum critical points awaiting experimental
findings. We will leave this potentially exciting explorations to
future studies.

Chao-Ming Jian's research at KITP is supported by
the Gordon and Betty Moore Foundations EPiQS Initiative through
Grant GBMF4304. Cenke Xu is
supported by the David and Lucile Packard Foundation.

\appendix

\section{Parton construction for the $N_f = 4$ QED$_3$}

\label{App:PartonQED}

\begin{figure}
\includegraphics[width=240pt]{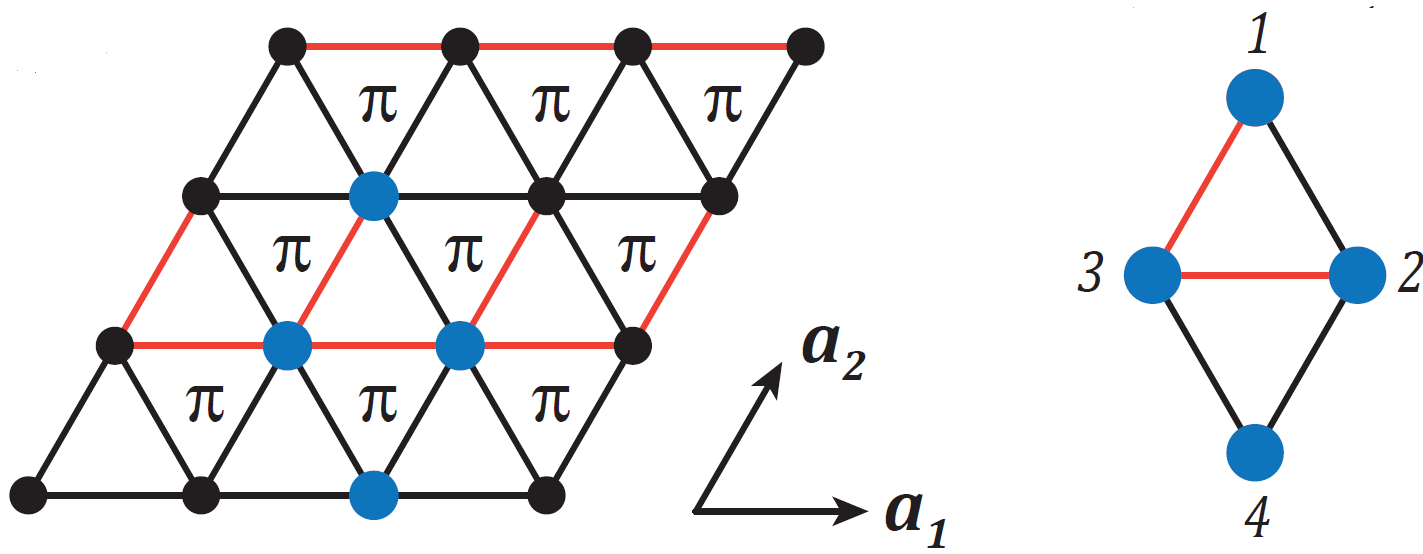}
\caption{(a) In the $\pi$-flux phase, we consider the mean-field
with only nearest-neighbor hopping. The hopping term takes value
$t$ along the black edges and $-t$ along the red edges. We choose
a convention such that the four blue sites form a unit cell. (b)
We provide an ordering of the sites within a unit cell for the
construction of the Bloch Hamiltonian. }
\label{Fig:Triangle_Pi_Flux}
\end{figure}

This section focuses on the parton construction, following
Ref.~\onlinecite{lutriangle}, of the $N_f = 4$ QED$_3$ for the
parton $f_s$ and $f_v$ discussed in Sec. \ref{Sec:With Hund's}.
Take the $f_s$ parton as an example. There are two fermion modes
$f_{s,\alpha}$ with $\alpha=1,2$ at each site. We impose the
constraint that there is exactly one $f_s$ fermion/parton per
site. In another words, this constraint enforces the half filling
of the $f_s$ fermions. In the following, we will suppress the
subscript ``$s$" to simplify the notation. In the following, we
will consider a mean-field ansatz that does not depend on the mode
index $\alpha=1,2$. Hence, we will also suppress the subscript
$\alpha$ for simplicity. The mean-field ansatz we consider
corresponds to the $\pi$-flux phase where we insert a $\pi$ flux
in every down triangle. In this $\pi$-flux phase, we only include
the nearest-neighbor hopping that takes value $t$ along the black
edges and $-t$ along the red edges as is shown in Fig.
\ref{Fig:Triangle_Pi_Flux} (a). A Bloch unit cell in this ansatz
contains 4 sites. We choose the convention such that the 4 blue
sites in Fig. \ref{Fig:Triangle_Pi_Flux} (a) form a unit cell. The
four fermions within a unit cell are organized into a 4-component
spinor according to the ordering shown in Fig.
\ref{Fig:Triangle_Pi_Flux} (b). The parton mean-field band
structure is given by the Bloch Hamiltonian
\begin{widetext}
\beqn
 H_{\rm mf}(k_1,k_2) =
t \left(
\begin{array}{cccc}
 0 & 1+e^{i k_2-i k_1} & -1+e^{i k_2} & e^{i k_2} \left(1+e^{-i k_1}\right) \\
 1+e^{i k_1-i k_2} & 0 & -1-e^{i k_1} & 1-e^{i k_2} \\
 -1+e^{-i k_2} & -1-e^{-i k_1} & 0 & 1+e^{i k_2-i k_1} \\
 e^{-i k_2} \left(1+e^{i k_1}\right) & 1-e^{-i k_2} & 1+e^{i k_1-i k_2} & 0 \\
\end{array}
\right), \label{Eq:mean-field Dirac} \eeqn
\end{widetext}
where $k_1$ and $k_2$ are defined by the phases $e^{i k_1 }$ and
$e^{i k_2 }$ obtained from translating the Bloch wave by one unit
cell along the $a_1$ and $a_2$ directions (see Fig.
\ref{Fig:Triangle_Pi_Flux}(a)). The translation $T_{1,2}$ by one
site along $a_1$ and $a_2$ directions need to be followed by gauge
transformations to keep the Hamiltonian $ H_{\rm mf}(k_1,k_2) $
invariant. The actions of $T_{1,2}$ on the 4-component spinor are
given by \beqn & U_{T_{1}} = \left(
\begin{array}{cccc}
 0 & 0 & 0 & e^{- i k_1+i k_2 } \\
 0 & 0 & 1 & 0 \\
 0 & e^{-i k_1} & 0 & 0 \\
 e^{-i k_2} & 0 & 0 & 0 \\
\end{array}
\right),
\nonumber \\
& U_{T_{2}} = i\left(
\begin{array}{cccc}
 0 & 0 & -i & 0 \\
 0 & 0 & 0 & -i \\
 i e^{-i k_2} & 0 & 0 & 0 \\
 0 & i e^{-i k_2} & 0 & 0 \\
\end{array}
\right). \eeqn
The mean-field band structure given by Eq. \ref{Eq:mean-field
Dirac} has a Dirac crossing at $(k_1, k_2) = (\pi,0)$ at half
filling, namely one $f$ fermion per-site. We can study the
low-energy physics at the Dirac crossing by considering $(k_1,
k_2) = (\pi + q_1, q_2)$ and expanding to the leading order of
$q_{1,2}$. We can also switch to the Cartesian coordinates $(q_x,
q_y)$ defined by the relation $(q_1 , q_2 ) = \left(q_x,
\frac{1}{2} q_x +  \frac{\sqrt{3}}{2} q_y\right)$. One can show
that, under a certain basis transformation $W$, the leading order
mean-field Bloch Hamiltonian reads: \beqn W^\dag H_{\rm mf} W  =
\sqrt{\frac{3}{2}}t \left( q_x \sigma^{0y}  + q_y \sigma^{0x}
\right) + ... , \eeqn where ``$...$" stands for terms that contain
higher powers of $q_x$ and $q_y$. Here, we've used the notation
$\sigma^{ab} \equiv \sigma^a \otimes \sigma^b$. $\sigma^0$
represents the $2\times 2$ identity matrix. We notice that the
leading order Hamiltonian $W^\dag H_{\rm mf} W$ describes two
copies of Dirac fermions that form together a doublet under an
emergent $SO(3)$ symmetry generated by $\sigma^{x0}, \sigma^{y0}$
and $\sigma^{z0}$. We can also extract the leading order terms of
the translation symmetry action $U_{T_{1,2}}$: \beqn W^\dag
U_{T_1} W= i \sigma^{z0},~~~~~ W^\dag U_{T_2} W= -i \sigma^{y0},
\eeqn which naturally corresponds to a $Z_2 \times Z_2$ subgroup
of the emergent $SO(3)$ symmetry. This emergent $SO(3)$ symmetry
is exactly the $SO(3)_m$ group discussed in the main text. The
low-energy Dirac fermion forms a doublet under $SO(3)_m$. Remember
we have suppressed the mode index $\alpha=1,2$ throughout the
discussion. Hence, there are in fact in total 4 copies of Dirac
fermions. These Dirac fermions also couple to a dynamical $U(1)$
gauge field that enforces the constraint on one $f$ fermion per
site. Hence, the $\pi$-flux phase can be described by the $N_f=4$
QED$_3$.

\bibliography{Moire_LSM}

\end{document}